\documentclass[twocolumn,pra,aps,floatfix,longbibliography]{revtex4-1}
\usepackage{amsmath,esint}
\usepackage{amsfonts}
\usepackage{amssymb}
\usepackage{gensymb}
\usepackage{siunitx}
\usepackage{verbatim}
\usepackage{graphicx}
\usepackage{xfrac}
\usepackage{times}
\usepackage{natbib}
\usepackage{hyperref}
\usepackage{dsfont}
\usepackage{url}
\usepackage{siunitx}
\usepackage{appendix}
\usepackage{multirow}
\usepackage{tabularx}
\begin{document}

\title{Viscous Maxwell-Chern-Simons theory for topological electromagnetic phases of matter}

\author{Todd Van Mechelen}
\affiliation{Purdue University, School of Electrical and Computer Engineering, Brick Nanotechnology Center, 47907, West Lafayette, Indiana, USA}
\author{Zubin Jacob}
\email{zjacob@purdue.edu}
\affiliation{Purdue University, School of Electrical and Computer Engineering, Brick Nanotechnology Center, 47907, West Lafayette, Indiana, USA}

\begin{abstract}

We present the fundamental model of a topological electromagnetic phase of matter: viscous Maxwell-Chern-Simons theory. Our model applies to a quantum Hall fluids with viscosity. We solve both continuum and lattice regularized systems to demonstrate that this is the minimal (exactly solvable) gauge theory with a nontrivial photonic Chern number ($C\neq 0$) for electromagnetic waves coupled to a quantum Hall fluid. The interplay of symmetry and topology is also captured by the spin-1 representations of a photonic skyrmion at high-symmetry points in the Brillouin zone. To rigorously analyze the topological physics, we introduce the viscous Maxwell-Chern-Simons Lagrangian and derive the equations of motion, as well as the boundary conditions, from the principle of least action. We discover topologically-protected chiral (unidirectional) edge states which minimize the surface variation and correspond to massless photonic excitations costing an infinitesimal amount of energy. Physically, our predicted electromagnetic phases are connected to a dynamical photonic mass in the integer quantum Hall fluid. This arises from viscous (nonlocal) Hall conductivity and we identify the nonlocal Chern-Simons coupling with the Hall viscosity. The electromagnetic phase is topologically nontrivial $C\neq 0$ when the Hall viscosity inhibits the total bulk Hall response. Our work bridges the gap between electromagnetic and condensed matter topological physics while also demonstrating the central role of spin-1 quantization in nontrivial photonic phases.

\end{abstract}

\maketitle 

\section{Introduction}

Chern-Simons theory has been studied in condensed matter and high-energy physics for over three decades \cite{Deser1982,DESER1982372,Jackiw1990,Dunne:1998qy}. In a two-dimensional (2D) quantum fluid, it describes the transverse current generated by an applied electric field, which manifests in the Hall conductivity $\sigma_{xy}$. Interestingly, 2D Chern-Simons theory also provides an elegant explanation of Hall quantization \cite{Klitzing1986} as well as the chiral edge currents, with no need to invoke electronic band structure. In addition, it has successfully described the fractional quantum Hall effect in many-body systems \cite{Tsui1982,Laughlin1983} and even enters the physics of anyons \cite{MOORE1991362,Wilczek1982}. On the other hand, three-dimensional (3D) Chern-Simons theory, also known as axion electrodynamics \cite{Wilczek1987}, emerges as a residual magnetoelectric response in topological insulators \cite{Moore2007,Qi2008,Wu1124}. The axion magnetoelectric effect accounts for the metallic surface current at the boundary of these topological insulating materials \cite{Qi1184}.

However, in both 2D and 3D, Chern-Simons theory only elucidates the topological manifestations of the electron; the topology of the electromagnetic field in these exotic materials has remained largely unexplored. Here, we mean quantities such as the \textit{photonic} Chern number and the topological invariants associated with the electromagnetic field coupled to condensed matter. To characterize these topological properties, it is fundamentally necessary to understand the dynamical $\omega\neq 0$ and subwavelength $k\neq 0$ behavior of the material response \cite{QuantGyro2018,VanMechelen:19}. In solids, this is encapsulated in the spatiotemporal dispersion of optical coefficients like the conductivity tensor $\sigma_{ij}(\omega,\mathbf{k})$, which cannot be explained by conventional Chern-Simons theory. This insight has led to a new electromagnetic classification of topological matter \cite{Nonlocal2019} and intriguing phenomena such as unidirectional electromagnetic spin waves \cite{Todd2019}. These so-called topological electromagnetic phases of matter are intrinsically bosonic (spin-1) and are fundamentally different from fermionic (spin-\sfrac{1}{2}) phases as they obey differing symmetries (e.g. time-reversal: $\mathcal{T}^2=+1$ for bosons vs. $\mathcal{T}^2=-1$ for fermions). The prototypical model of a gapped topological electromagnetic phase, with nontrivial photonic Chern number $C\neq 0$, was first connected to nonlocality (momentum dependence) of the Hall conductivity $\sigma_{xy}(k)=\kappa-\xi k^2$ \cite{QuantGyro2018}. In two dimensions, $\sigma_{xy}(k)$ transforms as a dynamical photonic mass and the topology of this system is interpreted as a spin-1 photonic skyrmion \cite{VanMechelen:19}. The goal of this paper is to rigorously analyze the associated nonlocal field theory and understand its physical origin - Hall viscosity. As such, we dub the problem as viscous Maxwell-Chern-Simons (MCS) theory.

\begin{figure}
    \centering
    \includegraphics[width=\linewidth]{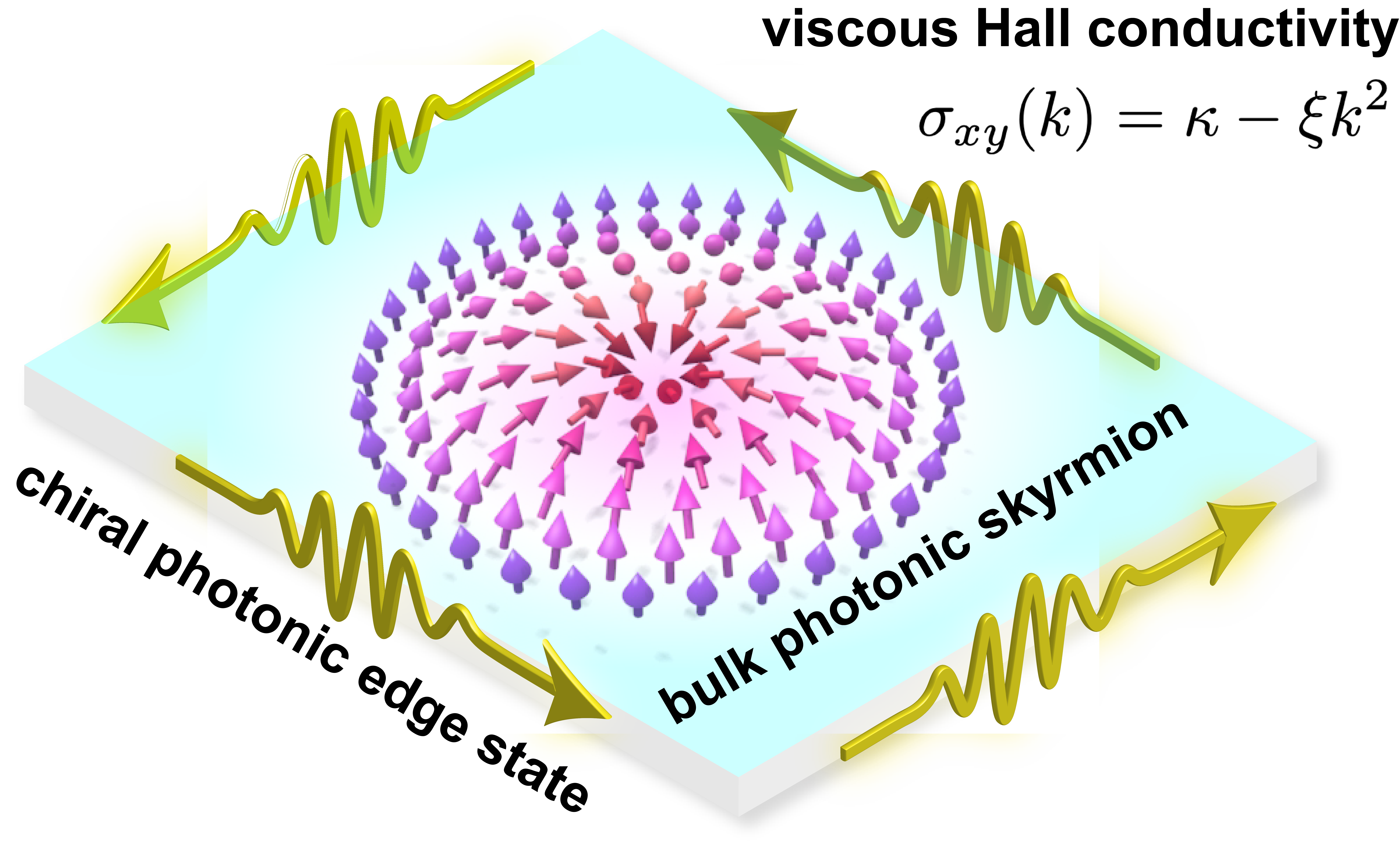}
    \caption{Overview of viscous Maxwell-Chern-Simons theory. The bulk topology is governed by a spin-1 photonic skyrmion which arises from viscous Hall conductivity $\sigma_{xy}(k)=\kappa-\xi k^2$. The boundary of the nontrivial phase $\kappa\xi>0$ hosts topologically-protected chiral photons which are linearly dispersing (massless).}
    \label{fig:Schematic}
\end{figure}

Hall viscosity $\eta_H$ \cite{Zaletel2013,Hughes2015,Moore2017}, also known as odd viscosity in fluid dynamics \cite{avron_odd_1998}, is a fundamental property of quantum Hall fluids and can exhibit topological quantization analogous to the Hall conductivity \cite{Avrom1995,Wen1992,Hoyos2012,Read2011}. Like conventional viscosity, it is related to the stress response of the system under deformations and governs the diffusive flow of the fluid. However, Hall viscosity is unique because it is dissipationless, inducing diffusive flow in a direction perpendicular to a pressure (force) gradient \cite{Sriram2017,Banerjee2017} and therefore does no work. In bulk momentum space, the Hall viscosity $\eta_H$ is expressed as a second order spatial correction to the Hall response,
\begin{equation}\label{eq:HallViscosity}
\frac{\sigma_{xy}(k)}{\sigma_{xy}(0)}=1-\frac{\eta_H}{\omega_c} k^2=1-D_H^2k^2.
\end{equation}
$\sigma_{xy}(0)$ is the intrinsic dc. Hall response and $D_H^2=\eta_H/\omega_c=\xi/\kappa$ is the Hall diffusion length, which is proportional to the Hall viscosity $\eta_H$. Depending on the material platform, Hall viscosity $\eta_H$ can be either positive $D_H^2>0$ or negative $D_H^2<0$ relative to the cyclotron motion $\omega_c=eB_0/(cm)$. This either inhibits or enhances the total Hall response. It has recently been shown that Hall viscosity which \textit{inhibits} the total response $D_H^2>0$ corresponds to a topologically nontrivial electromagnetic phase \cite{QuantGyro2018,VanMechelen:19,Nonlocal2019,Todd2019} and this has also been demonstrated for sound waves in fluid dynamics \cite{Souslov2019}. In an ideal quantum Hall fluid, $\sigma_{xy}(0)=Ne^2/(2\pi\hbar)$ is the quantum of conductance for the $N$th Landau level and $D_H^2=3Nl^2/2$ is on the order of the magnetic length $l=\sqrt{\hbar c/(eB_0)}$. $\hbar$ is the reduced Planck constant, $c$ is the speed of light, $e$ is the elementary charge and $B_0$ is the biasing magnetic field. Here, the Hall viscosity is not only quantized but also represents a topological phase for the photon $D_H^2>0$. An overview of the problem is depicted in Fig.~\ref{fig:Schematic}. Remarkably, Hall viscosity is no longer a theoretical proposition and has been measured experimentally in graphene's electron fluid \cite{Berdyugin162,Sherafati2016}; the viscous response is appreciable even under weak magnetic fields (classical fluid). Whats more, the measured viscosity is in the nontrivial regime $D_H^2>0$ and provides strong evidence of a topological electromagnetic phase of matter. In this paper, we present a general analysis of this topological phase and the unique phenomena associated with it. We have included a summary of a few physical systems exhibiting Hall viscosity along with their characteristic parameters in Tbl.~\ref{tab:Hall}.

For completeness, we note that our work is closely related to ideas in topological photonics but the physical platforms are fundamentally different as we are concerned with condensed matter systems. Topological photonics \cite{Lu2014,Khanikaev2017,Horsley2018,Ozawa2019} is a rapidly growing field of research where topological properties, i.e. quantities conserved under continuous deformations, are exploited for novel wave manipulation. These revelations in band theory, first discovered in condensed matter physics \cite{HasanKane10_RMP,Zhang2011,Po2017,Kruthoff2017,Farhad2019}, have transcended all of photonics; from plasmonics \cite{VanMechelen:16,Shi2018,Francesco2019}, metamaterials \cite{Lineaat2774,Shuang2019,Lustig2019} and photonic crystals \cite{Noh2018,Shalaev_2018,Yang2019}. Nevertheless, it remains an open question whether topological photonic phases can be expressed in terms of an effective gauge theory. A U(1) theory is fundamentally necessary if one attempts to add sources $J_\mu$ or canonically quantize the field. In both cases, we must work with the gauge fields $A_\mu$ themselves which requires a rigorous field theory rhetoric. Answering this question through viscous MCS theory is another important goal of the current work. 


\begin{table*}[htbp]
\caption{Summary of two physical systems exhibiting significant Hall viscosity $\eta_H$ and topologically nontrivial electromagnetic phases $C\neq 0$. In general, Hall viscosity is always present if the system breaks both parity and time-reversal symmetry. When Hall viscosity opposes the cyclotron motion $\eta_H/\omega_c>0$, the electromagnetic phase is nontrivial [Eq.~(\ref{eq:HallViscosity})], which occurs in both quantum Hall and graphene fluids. The quantum Hall fluid is an idealized zero temperature platform $T\to 0~\mathrm{K}$ at very large $B_0\gtrapprox  5 ~\mathrm{T}$. Remarkably, Hall viscosity is also appreciable in the graphene fluid \cite{Berdyugin162} around room temperature $100-300~\mathrm{K}$ and for weak magnetic fields $B_0\approx 10~\mathrm{mT}$.}
\label{tab:Hall}
\begin{tabular}{lll}
\hline\hline
 & quantum Hall fluid ($N=1$) & graphene fluid \cite{Berdyugin162}\\ \hline
Biasing field, $B_0$ &  $12 ~\mathrm{T}$ & $ 40~\mathrm{mT}$ \\
dc. Hall conductivity, $\sigma_{xy}(0)$ &  $Ne^2/(2\pi\hbar)=3.87\times 10^{-5}~\mathrm{S} $ & $8.01\times 10^{-2}~\mathrm{S}$ \\
Hall viscosity, $\eta_H$ & $3N\hbar/(2m)=1.74\times 10^{-4}~\mathrm{m^2/s}$ & $6.5\times 10^{-2}~\mathrm{m^2/s}$ \\
Cyclotron frequency, $\omega_c/(2\pi)$ & $336 ~\mathrm{GHz}~(1.07 ~\mathrm{mm})$ & $1.12~\mathrm{GHz} ~(267 ~\mathrm{mm})$ \\
Hall diffusion length, $D_H=\sqrt{\eta_H/\omega_c}$ & $\sqrt{3N\hbar c/(2eB_0)}=8.24~\mathrm{nm}$ &  $3.04 ~\mu\mathrm{m}$\\
Topological electromagnetic phase? & yes: $\eta_H/\omega_c>0$ & yes: $\eta_H/\omega_c>0$\\    \hline\hline         
\end{tabular}
\end{table*}

\section{Lagrangian formulation for topological electromagnetic phases}

\subsection{Maxwell-Chern-Simons theory}

We start with a brief review of conventional MCS theory. In 2+1 dimensions, the MCS Lagrangian is defined as \cite{Deser1982,DESER1982372,Jackiw1990,Dunne:1998qy},
\begin{equation}\label{eq:Lagrangian}
\mathcal{L}_A=-\frac{1}{4}F^{\mu\nu}F_{\mu\nu}-\frac{\kappa}{2}\epsilon^{\mu\nu\rho} A_\mu\partial_\nu A_\rho-A_\mu J^\mu.
\end{equation}
$A_\mu=(\phi,A_x,A_y)$ are the two-dimensional (2D) gauge fields and $F_{\mu\nu}=\partial_\mu A_\nu-\partial_\nu A_\mu$ is the field strength tensor. $J_\mu=(\rho,J_x,J_y)$ is a 2D conserved current $\partial_\mu J^\mu=0$. We have set the dielectric permittivity to unity $\varepsilon=1$ but the case with $\varepsilon>1$ is easily handled and does not alter the topological physics - it simply scales the field and the effective speed of light. The first term in $\mathcal{L}_A$ is the familiar Maxwell Lagrangian. The second term is the Chern-Simons Lagrangian and $\kappa$ is the coupling constant. Alternately, the MCS theory can be formulated in the more aesthetically pleasing ``self-dual" picture \cite{TOWNSEND198438,Diamantini1993},
\begin{equation}\label{eq:DualLagrangian}
\mathcal{L}_F=\frac{\kappa}{2}\Tilde{F}^\mu\Tilde{F}_\mu+\frac{1}{2}\epsilon^{\mu\nu\rho} \Tilde{F}_\mu\partial_\nu\Tilde{F}_\rho-\tilde{F}_\mu J^\mu,
\end{equation}
which is equivalent to Eq.~(\ref{eq:Lagrangian}) up to a Legendre transformation \cite{DESER1984371,FRADKIN198531}. In this case, the field theory is described in terms of the electromagnetic dual $\Tilde{F}^\mu=\frac{1}{2}\epsilon^{\mu\nu\rho}F_{\nu\rho}$, which satisfies the Bianchi identity (Faraday equation) $\partial_\mu\Tilde{F}^\mu=0$ upon variation of the action. In 2D, the dual field is a covariant vector $\tilde{F}_\mu=(B_z,E_y,-E_x)$ with the same number of components as the gauge fields $A_\mu=(\phi,A_x,A_y)$. This is why one can construct equivalent theories in terms of $A_\mu$ or $\tilde{F}_\mu$. Indeed, Lagrangians $\mathcal{L}_A$ and $\mathcal{L}_F$ produce the exact same equations of motion when one varies the action with respect to $A_\mu$ or $\tilde{F}_\mu$. The subtle difference is that $\mathcal{L}_F$ is manifestly gauge invariant, while $\mathcal{L}_A$ is only gauge invariant up to a total divergence. Another interesting feature is that the traditional self-dual theory contains no kinetic terms \cite{classical_1975}. As we will see, the inclusion of a kinetic term makes the field topologically nontrivial.

\subsection{Viscous Hall conductivity}

Although MCS theory has been studied extensively; only recently has nonlocality \cite{Fetter1985,Zhong2015} been considered and its topological implications on the electromagnetic field \cite{QuantGyro2018,VanMechelen:19,Nonlocal2019,Todd2019}. In the nonlocal theory, the Chern-Simons coupling $\kappa$ is spatially dispersive,
\begin{equation}
\kappa\to\int d\mathbf{r}'\Lambda(|\mathbf{r}-\mathbf{r}'|).
\end{equation}
The MCS theory is no longer relativistically invariant and cannot exist in vacuum. Nevertheless, nonlocality is always present within a material and will affect the electromagnetic response to varying degrees, depending on the system in question \cite{agranovich2013crystal,landau2013electrodynamics}. Physically, the Chern-Simons coupling is interpreted as a dissipationless Hall conductivity, as the induced current density is,
\begin{equation}\label{eq:HallCond}
J^\mu_\mathrm{ind}(\mathbf{r})=-\frac{\partial \mathcal{L}_A}{\partial A_\mu}-J^\mu=\int d\mathbf{r}'\Lambda(|\mathbf{r}-\mathbf{r}'|)\tilde{F}^\mu(\mathbf{r}').
\end{equation}
Since the induced current $J^\mu_\mathrm{ind}$ is proportional to the dual field $\tilde{F}^\mu$, the nonlocal conductivity tensor $\sigma_{ij}(|\mathbf{r}-\mathbf{r}'|)$ is,
\begin{equation}
\sigma_{xy}(|\mathbf{r}-\mathbf{r}'|)=-\sigma_{yx}(|\mathbf{r}-\mathbf{r}'|)=\Lambda(|\mathbf{r}-\mathbf{r}'|),
\end{equation}
with $\sigma_{xx}=\sigma_{yy}=0$. For our purposes, we need only consider the first correction to $\Lambda$,
\begin{equation}
\Lambda(|\mathbf{r}-\mathbf{r}'|)=\kappa\delta(\mathbf{r}-\mathbf{r}')+\xi\nabla^2\delta(\mathbf{r}-\mathbf{r}')+O(\nabla^4).
\end{equation}
We call $\xi$ the nonlocal Chern-Simons coupling. This term is sufficient to understand all the topological physics in the continuum limit $k\approx 0$, as the momentum space is now effectively $\mathbb{R}^2\simeq S^2$ \cite{Ryu_2010}. We also show in the lattice theory [Sec.~\ref{sec:Lattice}], that nontrivial photonic phases $C\neq 0$ are only possible when $\xi\neq 0$. By direct comparison with Eq.~(\ref{eq:HallViscosity}), we see that the nonlocal Chern-Simons coupling $\xi$ is proportional to the Hall viscosity $\xi/\kappa=D_H^2=\eta_H/\omega_c$ and governs the Hall diffusion. A forthcoming paper will present the topological electrohydrodynamics of viscous Hall fluids in more detail. Here, we only consider the low energy theory of an idealized quantum Hall fluid to demonstrate the importance of viscous (nonlocal) Hall conductivity.

\subsection{Viscous Maxwell-Chern-Simons theory}

Originally, Hall viscosity was conceived from a geometric perspective, associated with deformations of the metric \cite{Avrom1995}. An equivalent but simpler point of view is to include nonlocal terms that account for the stress-strain response. Therefore, we introduce the viscous MCS Lagrangian to elucidate this topological electromagnetic phase of matter,
\begin{equation}\label{eq:LagNonlocal}
\begin{split}
\mathcal{L}_A=-\frac{1}{4}F^{\mu\nu}F_{\mu\nu}&-\frac{\kappa}{2}\epsilon^{\mu\nu\rho} A_\mu\partial_\nu A_\rho\\ &+\frac{\xi}{2}\epsilon^{\mu\nu\rho} \nabla A_\mu\cdot\nabla\partial_\nu A_\rho,
\end{split}
\end{equation}
and we have set $J_\mu=0$ to analyze the free field theory. The physical meaning of the nonlocal term is not immediately obvious as we now have higher (cubic) derivatives in the Lagrangian \cite{Simon1990}. However, it has an intuitive interpretation in the self-dual picture,
\begin{equation}\label{eq:LagDualNonlocal}
\mathcal{L}_F=\frac{\kappa}{2}\Tilde{F}^\mu\Tilde{F}_\mu-\frac{\xi}{2}\nabla\Tilde{F}^\mu\cdot\nabla\Tilde{F}_\mu+\frac{1}{2}\epsilon^{\mu\nu\rho} \Tilde{F}_\mu\partial_\nu\Tilde{F}_\rho.
\end{equation}
$\xi\neq 0$ corresponds to a kinetic term and resembles the kinetic part of the Schr\"{o}dinger Lagrangian \cite{Arsenović2014}. Indeed, there is a one-to-one correspondence between Eq.~(\ref{eq:LagDualNonlocal}) and the minimal Dirac model \cite{SHEN2011},
\begin{equation}\label{eq:DiracLag}
\mathcal{L}_\psi=m\bar{\psi}\psi-B\nabla\bar{\psi}\cdot\nabla\psi-i\bar{\psi}\gamma^\mu\partial_\mu\psi,
\end{equation}
where $\gamma^\mu$ are the 2+1D gamma matrices and $\psi$ is a two-component spinor. Equations (\ref{eq:LagDualNonlocal}) and (\ref{eq:DiracLag}) are in fact supersymmetric partners \cite{DESER1982372}, describing spin-1 bosons and spin-\sfrac{1}{2} fermions respectively. By direct comparison, we see that $\kappa$ plays the role of photonic mass, in the same way as $m$ for the electron. Likewise, $\xi$ and $B$ dictate the kinetic (diffusive) terms, which are essential to realize nontrivial phases \cite{bernevig2013topological}. These terms add viscosity to the mass.

Again, $\mathcal{L}_A$ and $\mathcal{L}_F$ generate the same equations of motion when one varies the action with respect to $A_\mu$ or $\tilde{F}_\mu$. However, to ensure the action does not break gauge invariance on a boundary, it is more convenient to work with the self-dual theory $\mathcal{L}_F$. Varying the dual field $\tilde{F}_\mu\to\tilde{F}_\mu+\delta\tilde{F}_\mu$, we naturally obtain a bulk and surface term $\delta\mathcal{S}=\delta\mathcal{S}_b+\delta\mathcal{S}_s$,
\begin{equation}
\delta\mathcal{S}_b=\int dV \left[\partial_\mu F^{\mu\nu}+\left(\kappa+\xi\nabla^2\right)\tilde{F}^\nu\right]\delta\tilde{F}_\nu,
\end{equation}
and,
\begin{equation}
\delta\mathcal{S}_s=\int_{\partial V}dtdy\left[\left(\frac{1}{2}F^{x\mu}-\xi\partial_x \tilde{F}^\mu\right)\delta\tilde{F}_\mu\right]_{x=0}.
\end{equation}
$dV=dtdxdy$ is the differential space-time volume and we have taken the boundary at $x=0$. Clearly the bulk term vanishes $\delta \mathcal{S}_b=0$ if the wave equation is satisfied,
\begin{equation}\label{eq:EOM}
\partial_\mu F^{\mu\nu} +\left(\kappa+\xi\nabla^2\right) \Tilde{F}^\nu=0,
\end{equation}
where $F^{\mu\nu}=\epsilon^{\mu\nu\rho}\tilde{F}_\rho$ is the field strength. Equation~(\ref{eq:EOM}) represents the equations of motion of the viscous MCS theory. On the other hand, the surface term $\delta \mathcal{S}_s=0$ vanishes for two distinct boundary conditions. The first is a Dirichlet condition,
\begin{equation}\label{eq:OpenBC}
\delta\tilde{F}^\mu|_{x=0}=0,
\end{equation}
where the value of the field is fixed on $x=0$, usually to zero $\tilde{F}^\mu|_{x=0}=0$, corresponding to an open boundary. The second possibility is slightly more interesting and represents the natural (mixed) boundary condition,
\begin{equation}\label{eq:MixedBoudary}
\left[F^{x\mu}-2\xi\partial_x \tilde{F}^\mu\right]_{x=0}=0.
\end{equation}
Equation~(\ref{eq:MixedBoudary}) has a particularly nice explanation - it implies the Poynting vector (energy flux) normal to the boundary $P_x|_{x=0}=0$ vanishes. This will be more evident in Sec.~\ref{sec:Hamiltonian} where we establish the Hamiltonian (Schr\"{o}dinger) picture.

\begin{figure}
    \centering
    \includegraphics[width=\linewidth]{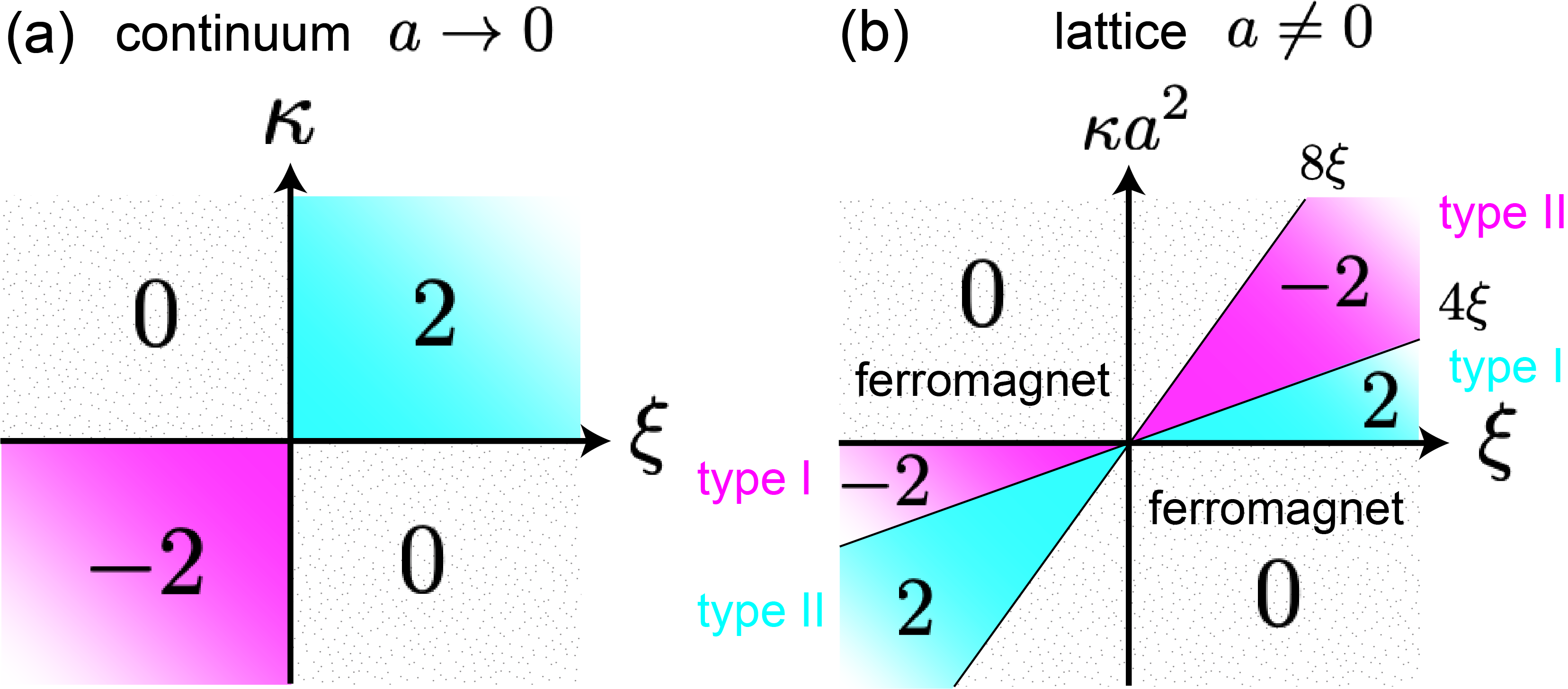}
    \caption{Topological phase diagrams for (a) continuum and (b) lattice models of viscous Maxwell-Chern-Simons theory. $C=\pm 2,0$ is the photonic Chern number of the positive energy band $\omega>0$ for different parameters. $\kappa$ and $\xi$ are the Chern-Simons and nonlocal Chern-Simons coupling respectively. $a$ is the lattice constant of a square grid. $\kappa a^2=0,4\xi,8\xi$
    denote the phase transition lines in the lattice model. These correspond to points of accidental degeneracy, where the band gap closes at $\mathbf{k}=\Gamma,X/ Y,M$ respectively. Importantly, conventional MCS theory $\xi=0$ always corresponds to a topologically trivial phase $C=0$ in the lattice regularization.}
    \label{fig:PhaseDiagram}
\end{figure}

\section{Hamiltonian (Schr\"{o}dinger) formulation}
\label{sec:Hamiltonian}

\subsection{Riemann-Silberstein vector}

Here, we introduce a quantum mechanical (Hamiltonian) formalism of the electromagnetic field. This procedure is incredibly useful to study the bulk and edge physics while also drawing direct parallels to condensed matter. To construct the electromagnetic ``Schr\"{o}dinger equation" it is very convenient to utilize the Riemann-Silberstein (RS) vector \cite{Bialynicki_Birula_2013}. In 2+1D, the RS vector is defined as,
\begin{equation}
\vec{F}=\begin{bmatrix}E_x & E_y & iB_z\end{bmatrix}.
\end{equation}
In the Schr\"{o}dinger picture $\vec{F}$ is treated as a 3D vector propagating in the 2D plane, while $\tilde{F}^\mu$ is a covariant vector, but the two are equivalent up to a unitary transformation. We now combine Eq.~(\ref{eq:EOM}) with the Bianchi identity (Faraday equation) to obtain a first-order wave equation,
\begin{equation}\label{eq:EMShrodinger}
i\partial_t \vec{F}=i\vec{d}\times\vec{F}=H\vec{F}.
\end{equation}
$\vec{d}$ is a 3D vector operator,
\begin{equation}
\vec{d}=\begin{bmatrix}p_x & p_y & \kappa-\xi p^2 \end{bmatrix},
\end{equation}
and $p_j=-i\partial_j$ are the corresponding momentum operators in the plane. $H$ is the ``Maxwell Hamiltonian" and can be expressed as the projection of $\vec{d}$ onto the vector spin operators $\vec{S}=\begin{bmatrix} S_x & S_y & S_z \end{bmatrix}$,
\begin{equation}\label{eq:ElectroHamil}
H=\vec{d}\cdot\vec{S}=p_xS_x+p_yS_y+(\kappa-\xi p^2)S_z.
\end{equation}
Up to a unitary transformation, Eq.~(\ref{eq:ElectroHamil}) is identical to the Navier-Stokes equations of a compressible time-reversal and parity breaking 2D fluid \cite{Souslov2019}; $\xi$ is the odd viscosity. In the RS basis, $[S_j,S_k]=i\epsilon_{jkl}S^l$ are antisymmetric SO(3) matrices that generate the spin-1 algebra \cite{Zhu2017,Tan2018,Fulga2018}. For time-dependent fields $\omega\neq 0$, Eq.~(\ref{eq:ElectroHamil}) automatically enforces Gauss's law from the definition of the cross product,
\begin{equation}
\vec{d}\cdot\vec{F}=\partial_iE^i-(\kappa+\xi \nabla^2)B_z=0.
\end{equation}
As we can see, Hall conductivity ties electric charge to the magnetic field $B_z$. However, an important difference between the Maxwell Hamiltonian ($H$) and Schr\"{o}dinger/Dirac Hamiltonians is that $\det[H]=0$ is manifestly singular (noninvertible). This is because the equations of motion for gauge theories are inherently redundant. Electrostatic $\omega=0$ (time-independent) potentials always exist $\vec{F}_0=\vec{d}\phi$ which are trivial solutions of $H\vec{F}_0=0$. Elimination of these ``zero modes" requires Gauss's law as an additional constraint at zero frequency $\vec{d}\cdot\vec{F}_0=\vec{d}\cdot\vec{d}\phi=[\nabla^2-(\kappa+\xi \nabla^2)^2]\phi=0$. These scalar potentials $\phi$ do not enter the free field theory since they do not possess a plane wave representation.

The boundary conditions have an intuitive interpretation in the RS basis. The open (Dirichlet) boundary condition [Eq.~(\ref{eq:OpenBC})] implies all components of the field vanish at $x=0$,
\begin{equation}\label{eq:OpenRS}
\vec{F}|_{x=0}=0.
\end{equation}
On the other hand, the natural boundary condition [Eq.~(\ref{eq:MixedBoudary})] guarantees that the normal Poynting vector $P_x|_{x=0}=0$ vanishes. To see this explicitly, it is useful to first derive the velocity operator $v_i$,
\begin{equation}
v_i=\frac{\partial H}{\partial p_i}=S_i-2\xi p_i S_z,
\end{equation}
which has an additional kinetic term when $\xi\neq 0$. The Poynting vector is simply the expectation value of $P_i=\vec{F}^*\cdot v_i\vec{F}$. Clearly, if $\vec{F}$ is a null state of $v_x$ at $x=0$,
\begin{equation}\label{eq:MixedRS}
v_x\vec{F}|_{x=0}=0,
\end{equation}
then the normal component $P_x|_{x=0}=0$ vanishes identically. Equation~(\ref{eq:MixedRS}) is equivalent to Eq.~(\ref{eq:MixedBoudary}), just expressed in a more enlightening form. 

\subsection{Dynamical photonic mass}

The topological physics is governed almost entirely by the Chern-Simons coupling - aka. the Hall conductivity $\Lambda$. To appreciate its significance, we translate the system to the energy-momentum space and analyze the bulk eigenstates of the theory. In the reciprocal space, $\vec{d}(\mathbf{k})=\begin{bmatrix} k_x & k_y & \Lambda(k) \end{bmatrix}$ is simply a vector and $\Lambda(k)=\kappa-\xi k^2$ is a function of the momentum. The Hamiltonian $H(\mathbf{k})=\vec{d}(\mathbf{k})\cdot\vec{S}$ now resembles the Zeeman interaction \cite{han2017skyrmions} where $\vec{d}$ plays the role of the magnetic field. The essential difference is that $\vec{S}$ are spin-1 (SO(3)) operators, as opposed to the Pauli matrices $\vec{\sigma}$ which are spin-\sfrac{1}{2} (SU(2)) operators. The dispersion relation of the dynamical $\omega\neq 0$ modes is found straightforwardly,
\begin{equation}
\omega^2(k)=\vec{d}^2(\mathbf{k})=k^2+\Lambda^2(k).
\end{equation}
The positive energy $\omega=d>0$ bulk eigenstate is then derived as,
\begin{equation}
\begin{split}
\vec{F}_\mathbf{k}&=\frac{1}{\sqrt{2}}\left[\frac{\vec{d}\times\hat{z}}{|\vec{d}\times\hat{z}|}+i\frac{\vec{d}\times(\vec{d}\times\hat{z})}{d|\vec{d}\times\hat{z}|}\right]\\
&=\frac{1}{\sqrt{2}}\left[-i\frac{\Lambda}{\omega}\hat{k}+\hat{\varphi}+i\frac{k}{\omega}\hat{z}\right],
\end{split}
\end{equation}
which has been normalized to unit energy $|\vec{F}_\mathbf{k}|^2=|\mathbf{E}|^2+|B_z|^2=1$. The Chern-Simons coupling (Hall conductivity) $\kappa\neq 0$ behaves as a gauge invariant photonic mass $\Lambda$ that opens a band gap at $\omega=0$. Note, the MCS mass should not be confused with the Proca mass which breaks gauge invariance. Although gauge invariant, the MCS mass does not preserve parity or time-reversal symmetry, admitting the possibility of nontrivial Chern phases $C\neq 0$. The Hall viscosity $\xi\neq 0$ makes this photonic mass spatially dispersive $\Lambda(k)=\kappa-\xi k^2$ and is crucial to realize these nontrivial phases.

\subsection{Photonic (spin-1) skyrmion}

Importantly, the MCS mass $\Lambda(k)$ also defines the representation theory of the 2+1D Poincar\'e algebra \cite{Dunne:1998qy},
\begin{equation}
j_m=\frac{\Lambda(k)}{|\Lambda(k)|}=\mathrm{sgn}[\Lambda(k)],
\end{equation}
which is a massive spin-1 excitation $j_m=\pm 1$. The representation $j_m$ indicates whether the plane wave is right ($+1$) or left ($-1$) circularly polarized in the $x$-$y$ plane. The topology is intimately tied to the spin-1 representation of the electromagnetic field. The Berry curvature $\Omega$ is precisely the ``magnetic field" of a spin-1 skyrmion \cite{VanMechelen:19},
\begin{equation}
\begin{split}
\Omega&=-i\left(\partial_x\vec{F}_\mathbf{k}^*\cdot\partial_y\vec{F}_\mathbf{k}-\partial_y\vec{F}_\mathbf{k}^*\cdot\partial_x\vec{F}_\mathbf{k}\right)\\
&=\hat{d}\cdot(\partial_x\hat{d}\times\partial_y\hat{d}).
\end{split}
\end{equation}
Spin-1 skyrmions are particularly interesting because the Chern number is always an even integer $C\in 2\mathbb{Z}$,
\begin{equation}
C=\frac{1}{2\pi}\int d\mathbf{k}~\Omega=2N. 
\end{equation}
$N\in \mathbb{Z}$ is the skyrmion winding number \cite{Nagaosa2013} that counts the number of times $\hat{d}$ wraps around the unit sphere,
\begin{equation}
N=\frac{1}{4\pi}\int d\mathbf{k}~\hat{d}\cdot(\partial_x\hat{d}\times\partial_y\hat{d}).
\end{equation}
Due to continuous rotational symmetry about $\hat{z}$, the Chern number is determined by $S_z$ eigenvalues at $k=0$ and $k=\infty$,
\begin{equation}
C=\hat{d}_z(0)-\hat{d}_z(\infty),
\end{equation}
which is exactly the difference in spin-1 representations $j_m$ at $k=0$ and $k=\infty$ respectively \cite{fang2017topological},
\begin{equation}\label{eq:ChernCont}
C=\mathrm{sgn}[\Lambda(0)]-\mathrm{sgn}[\Lambda(\infty)]=\mathrm{sgn}(\kappa)+\mathrm{sgn}(\xi).
\end{equation}
In the nontrivial regime $\kappa\xi>0$, the unit vector $\hat{d}$ wraps around the sphere once $|N|=1$ and corresponds to a Chern number of $|C|= 2$. In this case, the representation changes $\mathrm{sgn}[\Lambda(0)]\neq\mathrm{sgn}[\Lambda(\infty)]$ as there is point where the MCS mass passes through zero $\Lambda(k_c)=0$, precisely at $k_c=\sqrt{\kappa/\xi}$. Another important point; the quadratic spatial dispersion $\xi k^2$ naturally regularizes the continuum theory as the momentum space is equivalent to $\mathbb{R}^2\simeq S^2$ \cite{Ryu_2010}. This is what guarantees Chern number quantization \cite{QuantGyro2018,VanMechelen:19,Nonlocal2019,Todd2019}. As a visualization, the topological phase diagram of the continuum theory is displayed in Fig.~\ref{fig:PhaseDiagram}(a).

\begin{figure}
    \centering
    \includegraphics[width=\linewidth]{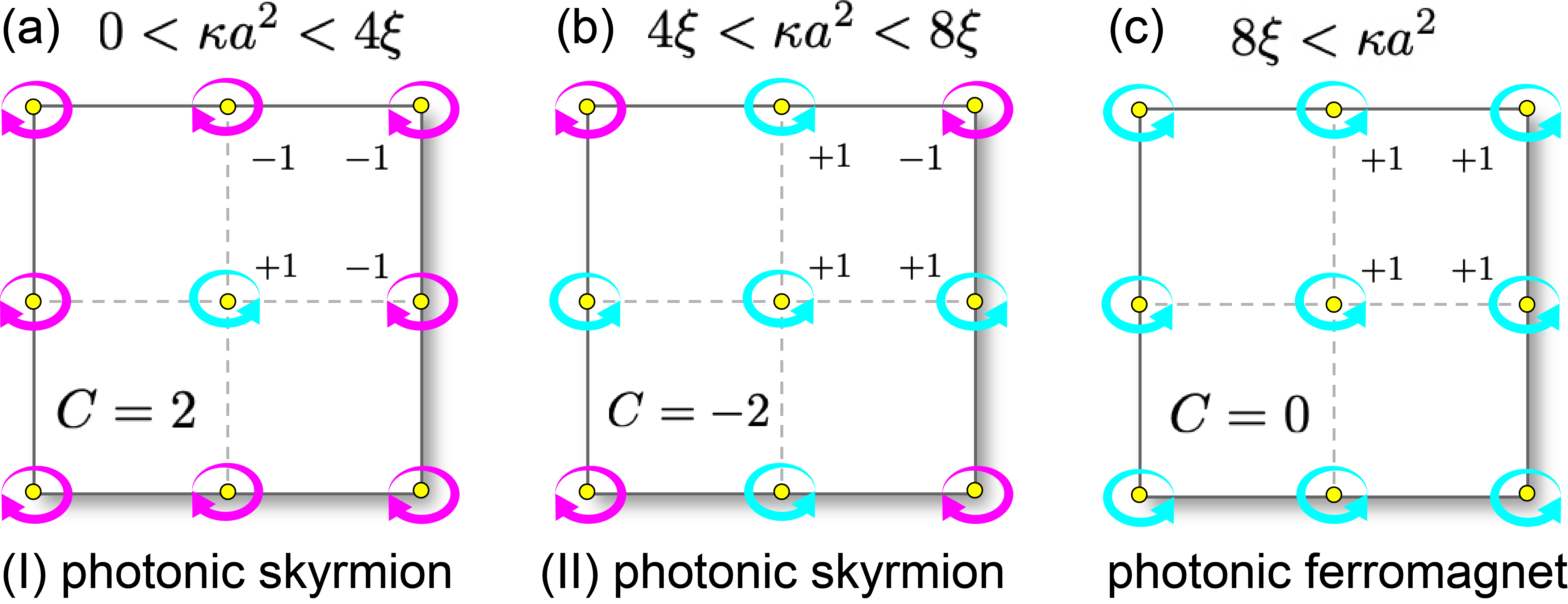}
    \caption{(a), (b) and (c) show the three phases $C=\pm 2,0$ of the lattice regularized theory. $\kappa>0$ and $\xi>0$ are chosen positive such that (a) and (b) label type I and type II photonic skyrmions respectively. (c) is the photonic ferromagnet. The eigenvalue at high-symmetry points denotes the sign of the Maxwell-Chern-Simons mass $j_m=\mathrm{sgn}(\Lambda)=\pm 1$, which determines the spin-1 representation; i.e. if the field is right ($+1$) or left ($-1$) circularly polarized. The two nontrivial phases possess skyrmion numbers of $N=\pm 1$ corresponding to a spin-1 Chern number of $C=2N=\pm 2$.}
    \label{fig:brillouin_zone}
\end{figure}

\section{Lattice formulation}\label{sec:Lattice}

\subsection{Bulk topology}

Finally, we place the MCS theory on a lattice \cite{Drell1979,Becchi1991,Chew1994,HOFERICHTER1995358,Teixeira1999,Teixeira2014} to demonstrate that Hall viscosity $\xi$, is essential for topological electromagnetic phases. Now space is discretized, $x=n_xa$ and $y=n_y a$, with $n_{x,y}\in\mathbb{Z}$. Here, time remains continuous and we assume a square lattice for simplicity; $a$ is the lattice spacing. First-order derivatives are converted to,
\begin{equation}
\partial_x\vec{F}\to \frac{\vec{F}(x+a)-\vec{F}(x-a)}{2a},
\end{equation}
while second-order derivatives are,
\begin{equation}
\partial_x^2\vec{F}\to \frac{\vec{F}(x+a)-2\vec{F}(x)+\vec{F}(x-a)}{a^2},
\end{equation}
with similar expressions for $y$. The continuum limit $a\to 0$ is obtained from standard calculus assuming the fields are at least twice differentiable \cite{weir2010thomas}. Transferring to the momentum space $\vec{F}=e^{i\mathbf{k}\cdot\mathbf{r}}\vec{F}_\mathbf{k}$; linear terms in the Hamiltonian [Eq.~(\ref{eq:ElectroHamil})] are replaced with,
\begin{equation}
p_x\vec{F}\to \frac{1}{a}\sin (k_x a)\vec{F}_\mathbf{k},
\end{equation}
and similarly for quadratic terms arising from the Hall viscosity $\xi\neq 0$,
\begin{equation}
p_x^2\vec{F}\to \frac{4}{a^2}\sin^2\left(\frac{k_x a}{2}\right) \vec{F}_\mathbf{k}.
\end{equation}
Due to discretization, the momentum is only unique up to $|k_{x,y}|\leq \pi/a$, which defines a torus $\mathbb{T}^2$ in two dimensions. The bulk dispersion relation is now given as,
\begin{equation}
\omega^2(\mathbf{k})=\vec{d}^2(\mathbf{k})=\frac{1}{a^2}[\sin^2(k_xa)+\sin^2(k_ya)]+\Lambda^2(\mathbf{k}),
\end{equation}
where $\vec{d}(\mathbf{k}+\mathbf{g})=\vec{d}(\mathbf{k})$ is periodic in the reciprocal lattice and $g_{x,y}=N_{x,y}2\pi/a$ is an arbitrary reciprocal vector $N_{x,y}\in\mathbb{Z}$,
\begin{equation}
\vec{d}(\mathbf{k})=\begin{bmatrix}
 a^{-1}\sin (k_x a) &  a^{-1}\sin (k_y a) & \Lambda(\mathbf{k})
\end{bmatrix}.
\end{equation}
$\Lambda(\mathbf{k})$ is the effective photon mass in the lattice theory,
\begin{equation}
\Lambda(\mathbf{k})=\kappa-\xi\left(\frac{2}{a}\right)^2\left[\sin^2\left(\frac{k_x a}{2}\right)+\sin^2\left(\frac{k_y a}{2}\right)\right].
\end{equation}
It is easy to check that the continuum limit [Eq.~(\ref{eq:HallCond})] is recovered when $a\to 0$.

\begin{figure}
    \centering
    \includegraphics[width=\linewidth]{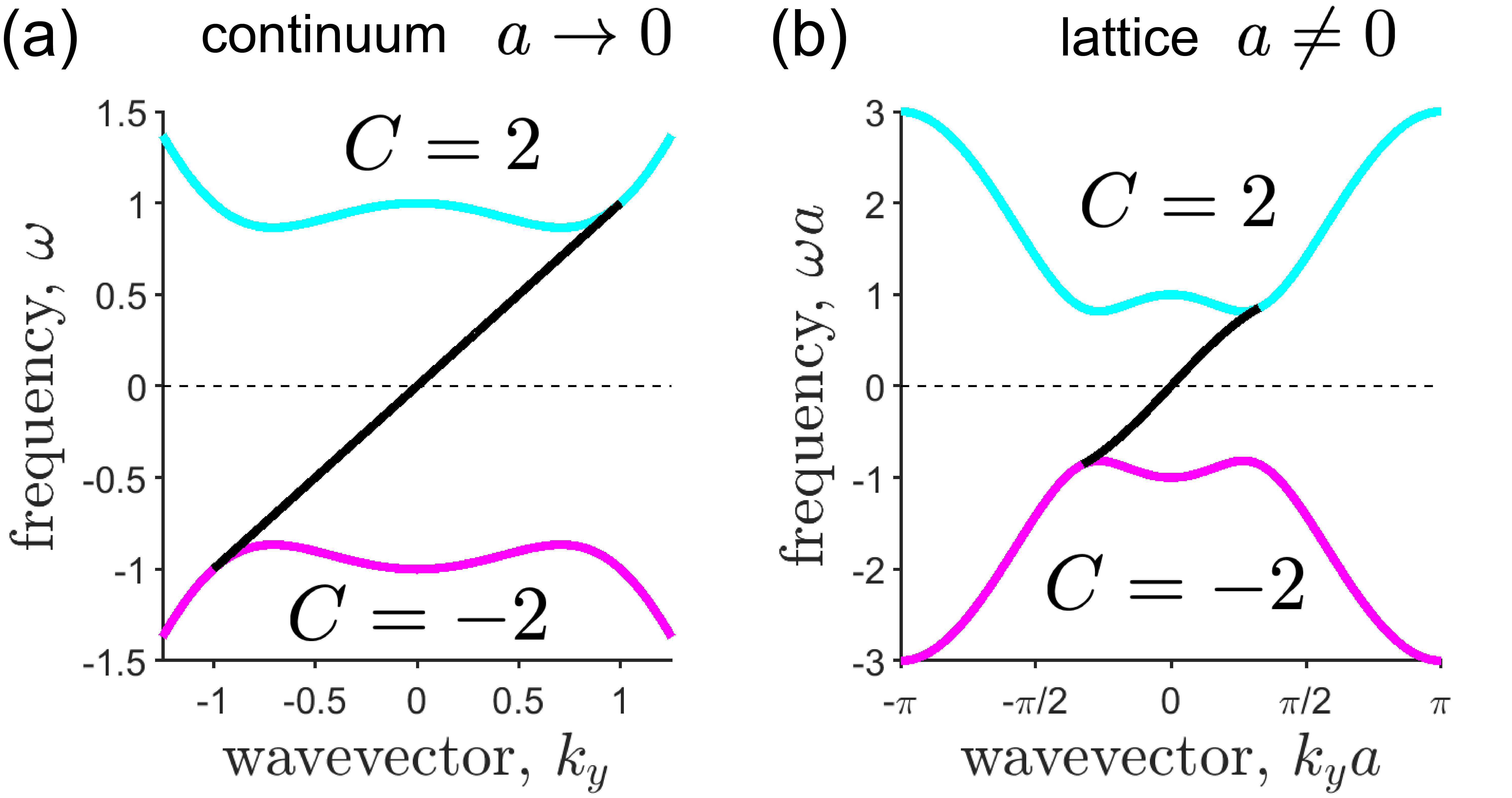}
    \caption{Bulk and edge dispersion of (a) continuum and (b) lattice models of viscous Maxwell-Chern-Simons theory. Cyan and magnetic lines are positive and negative energy topological bands while the black line is the chiral edge state. (a) Parameters are $\kappa=\xi=1$ in the continuum theory $a\to 0$. (b) Parameters are $\kappa a=\xi/a=1$ in the lattice theory $a\neq 0$.}
    \label{fig:dispersion}
\end{figure}

The importance of Hall viscosity $\xi$ is even more apparent in the lattice theory. At high-symmetry points $\mathbf{k}=\Gamma,X/Y,M$, the spin-1 representation \cite{Po2017,Kruthoff2017} only changes if $\xi\neq 0$. After a bit of work, it can be shown that the Chern number is \cite{Nascimento2017},
\begin{equation}\label{eq:LatticeChern}
\begin{split}
C&=\mathrm{sgn}[\Lambda(\Gamma)]+\mathrm{sgn}[\Lambda(M)]-2\mathrm{sgn}[\Lambda(X)]\\
&=\mathrm{sgn}(\kappa)+\mathrm{sgn}(\kappa-8\xi/a^2)-2\mathrm{sgn}(\kappa-4\xi/a^2).
\end{split}
\end{equation}
The eigenvalues at $\Lambda(X)=\Lambda(Y)$ are identical and thus appears twice in Eq.~(\ref{eq:LatticeChern}). For standard MCS theory $\xi=0$, the Chern number is identically zero $C=0$ in the lattice regularization. This is due to the inherent field doubling that occurs in a periodic system \cite{Goswani1997,OLESEN2015303} which cancels any parity anomalies that may arise in the continuum limit. Hence, $\kappa$ alone cannot describe a topological photonic phase. A nontrivial phase $|C|=2$ is only possible when $\xi\neq 0$. This is the spin-1 photonic analogue of the Haldane Chern insulator \cite{Haldane1988}, where the effective mass changes sign at high-symmetry points. It is also clear that we recover the continuum Chern number [Eq.~(\ref{eq:ChernCont})] in the limit of $a\to 0$ for $\xi\neq 0$. As a visualization, examples of type I and type II photonic skyrmions are displayed in Fig.~\ref{fig:brillouin_zone}. 

\begin{figure}
    \centering
    \includegraphics[width=\linewidth]{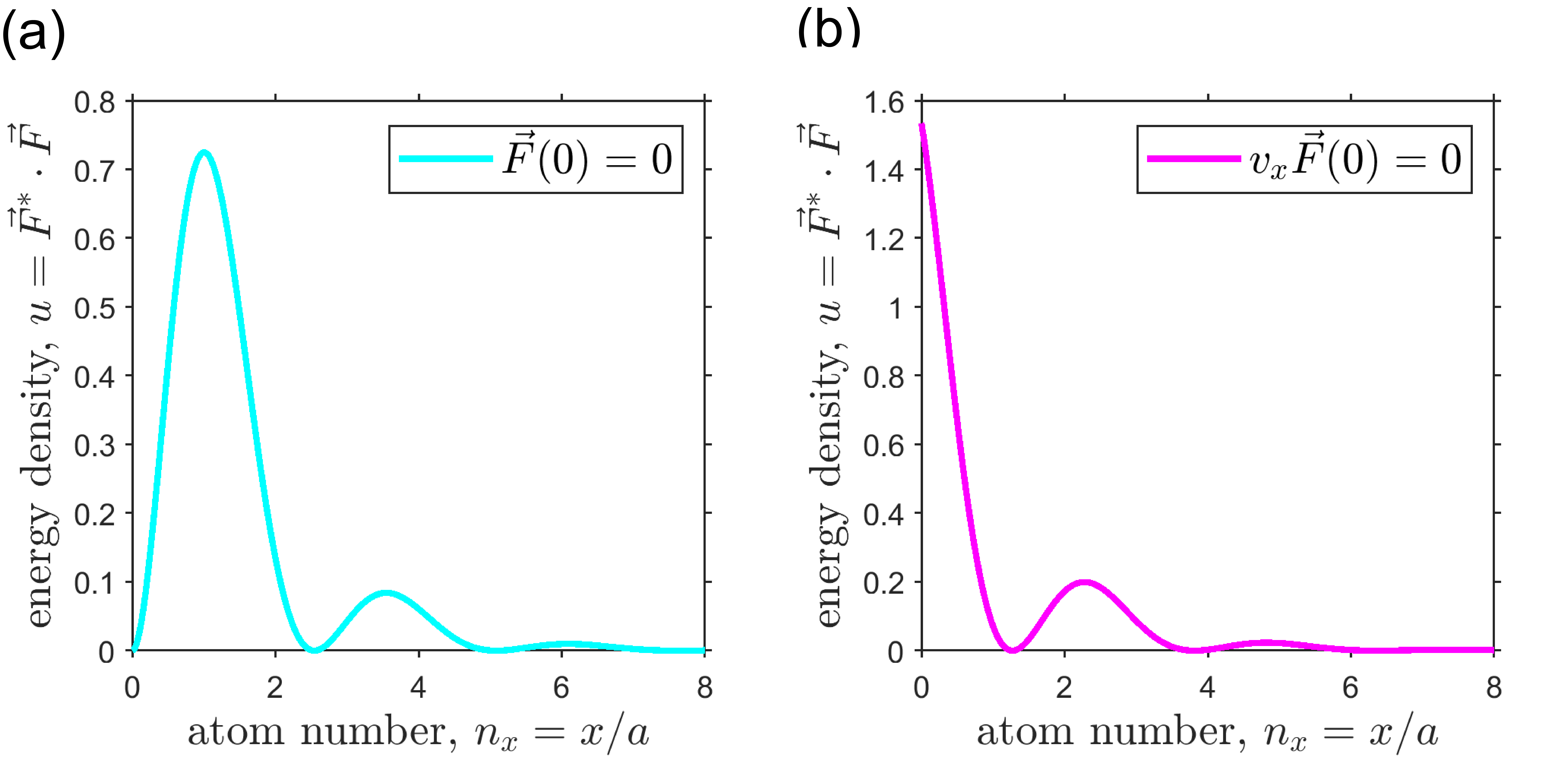}
    \caption{The two boundary conditions that minimize the surface variation $\delta\mathcal{S}_s=0$. (a) and (b) plot the normalized energy density $u=|\vec{F}|^2=|\mathbf{E}|^2+|B_z|^2$ of the chiral edge state. The parameters are $\kappa a=0.1$,
    $\xi/a=0.2$ and $k_ya=0.1$ as a demonstration. (a) The Dirichlet (open) boundary condition $\vec{F}(0)=0$ has zero measure at $x=0$. (b) The natural boundary condition $v_x\vec{F}(0)=0$ is more localized at the surface and resembles an evanescent wave.}
    \label{fig:boundary_conditions}
\end{figure}

\subsection{Protected spin-1 edge states}

Topologically-protected edge states in the lattice theory assume a nearly identical form as the continuum limit \cite{QuantGyro2018,VanMechelen:19}, with only slight modifications to the dispersion. The bulk-boundary correspondence (BBC) is also derived in a very similar manner. Instead of repeating the steps here, we will simply state the salient results; we have appended the rigorous BBC proof to the supplementary information. We strongly recommend consulting the paper by Mong (Ref.~\cite{Mong2011}) for an exhaustive derivation of the BBC which has been closely adapted here.

To uncover the edge states, we terminate the lattice at $x=0$ and introduce a half-space in the $x>0$ domain. In this case, $|k_y|\leq \pi/a$ is still a good quantum number, which we Fourier transform over, but the lattice has now been truncated at $x=n_x a=0$. Hence, we solve for the field at every discrete lattice point $n_x\geq 0$ which is labelled by $\vec{F}(n_x a)$ and the stipulation that the field is decaying $\vec{F}(n_x a)\to 0$ as $n_x\to \infty$. To satisfy an open $\vec{F}(0)=0$ or mixed $v_x\vec{F}(0)=0$ boundary condition, the edge state must possess a degenerate eigenvector for two decay constants $\eta_{1,2}$. After a bit of work, the edge state can be expressed as,
\begin{equation}
\vec{F}(n_x a)=\begin{bmatrix}
1\\ 0 \\ -s_C i
\end{bmatrix} \left[c_1\exp^{n_x}(-\eta_1 a )+c_2\exp^{n_x}(-\eta_2 a )\right],
\end{equation}
where $s_C=\mathrm{sgn}(C)$ is the sign of the Chern number and $c_{1,2}$ are proportionality constants that are set by the boundary condition. The dispersion relation reads,
\begin{equation}
\omega a= s_C\sin(k_ya).
\end{equation}
For small $k_ya\approx 0$ the dispersion is linear $\omega\approx s_C k_y$ and the edge state propagates near the speed of light $|\partial \omega/\partial k_y|=1$. The bulk and edge dispersion of the lattice model is shown in Fig.~\ref{fig:dispersion}. Interestingly, $S_y\vec{F}_{n_x}=s_C\vec{F}_{n_x}$ is also an eigenstate of the spin-1 helicity operator (SO(3)) with quantized spin along $\hat{y}$.

The proportionality constants $c_{1,2}$ can be chosen to satisfy Dirichlet (open) $\vec{F}(0)=0$ or mixed (natural) boundary conditions $v_x\vec{F}(0)=0$ at $x=n_xa=0$. The Dirichlet condition represents an antisymmetric combination $c_1=-c_2$ while the natural condition is a symmetric combination $c_1=c_2$. A depiction of the two boundary conditions is shown in Fig.~\ref{fig:boundary_conditions}. The decay constants $\eta_{1,2}$ are found from the characteristic equation, with $k_x=i\eta$,
\begin{equation}
\frac{s_C}{a}\sinh(\eta a)=\Lambda(\eta),
\end{equation}
which has two decaying roots $\Re(\eta_{1,2})>0$ strictly in the nontrivial regime $|C|=2$. Solving the secular equation we obtain,
\begin{equation}
\exp(-\eta_{1,2} a)=\frac{1}{2q_+}\left(-p\pm\sqrt{w^2-4q_+q_-}\right)
\end{equation}
where,
\begin{equation}
p=\kappa-\left(\frac{2}{a}\right)^2\xi\left[\sin^2\left(\frac{k_y a}{2}\right)+\frac{1}{2}\right], ~~~ q_\pm=\frac{s_C}{2a}\pm\frac{\xi}{a^2}.
\end{equation}
The regimes where $\Re(\eta)>0$ define the allowed parallel $k_y$ vectors of the edge state. When one of $\eta=0$, the edge state is no longer confined and this occurs when $k_y$ intersects the bulk bands; i.e. the edge state is gapless.

\section{Conclusions}

We have presented viscous Maxwell-Chern-Simons theory; the fundamental (exactly solvable) model of a topological electromagnetic phase. The topological physics is ultimately governed by viscous (nonlocal) Hall conductivity. To rigorously analyze the problem, we introduced the nonlocal Maxwell-Chern-Simons Lagrangian and derived the equations of motion, as well as the boundary conditions, from the principle of least action. A bulk-boundary correspondence was proven for the topologically-protected edge states and it was shown that they minimize the surface variation.

\section*{Acknowledgements}

This research was supported by the Defense Advanced Research Projects Agency (DARPA) Nascent Light-Matter Interactions (NLM) Program and the National Science Foundation (NSF) [Grant No. EFMA-1641101].

\bibliography{top_optics.bib}

\end{document}